\documentclass{Interspeech2024}

\interspeechcameraready

\title{Predicting Heart Activity from Speech using Data-driven and Knowledge-based features}

\name[affiliation={1,2}]{Gasser}{Elbanna}
\name[affiliation={2,3}]{Zohreh}{Mostaani}
\name[affiliation={2}]{Mathew}{Magimai.-Doss}

\address{
  $^1$Speech and Hearing Bioscience and Technology, Harvard Medical
School, MA, USA\\
  $^2$Idiap Research Institute, Martigny, Switzerland \\
  $^3$Ecole polytechnique f\'ed\'erale de Lausanne, Switzerland}
\email{gasser\_elbanna@hms.harvard.edu, \{zohreh.mostaani, mathew\}@idiap.ch}


\begin{document}

\maketitle

\begin{abstract}
    
    Accurately predicting heart activity and other biological signals is crucial for diagnosis and monitoring. Given that speech is an outcome of multiple physiological systems, a significant body of work studied the acoustic correlates of heart activity. Recently, self-supervised models have excelled in speech-related tasks compared to traditional acoustic methods. However, the robustness of data-driven representations in predicting heart activity remained unexplored. In this study, we demonstrate that self-supervised speech models outperform acoustic features in predicting heart activity parameters. We also emphasize the impact of individual variability on model generalizability. These findings underscore the value of data-driven representations in such tasks and the need for more speech-based physiological data to mitigate speaker-related challenges.
\end{abstract}

\section{Introduction}
\label{sec:intro}
In the contemporary era of telemedicine, harnessing ubiquitous biological signals such as human speech has been of great importance. Thanks to the advances in machine learning, a significant body of research has emerged, exploring paralinguistic analysis in speech. This multi-faceted discipline encompasses tasks such as detecting various stress indicators \cite{gallardoantolin19_interspeech,elbanna22_interspeech_short}, predicting emotional states \cite{scheidwasser2022serab_short}, and modeling physiological parameters such as heart activity \cite{schuller2013automatic_short}, respiration patterns \cite{nallanthighal2021deep_short}, and skin conductance responses \cite{schuller2014munich_short}.

There has been an effort to examine the association between speech signals and heart activity. Orlikoff et al. \cite{orlikoff1989effect} in one of the very early studies showed  that cardio vascular system can influence the vocal fundamental frequency (F0) indicating that the absolute F0 perturbation (jitter) during a sustained phonation could vary between 0.5\% to 20\%. In other studies, heart activity was studied in relation with speech in different emotional states. Williams et al. \cite{williams1972emotions} demonstrated that the emotion variation might cause an increase in blood pressure (BP), heart rate (HR), sub-glottal pressure, and the depth of respiratory movements. James et al. \cite{james2015heart} suggested a strong correlation between speech, emotion, and heart rate using spectral features from speech. Smith et al. \cite{smith2017analysis_short} showed that HR increases when the person is speaking compared to when they are silent and this increase is greater when they are frustrated. In another study \cite{ryskaliyev2016speech_short}, the HR was predicted using linear models in different emotional states. Jati et al. \cite{jati2018towards_short} predicted physiological signals from speech during stressful conversations.

However, the influence of inter- and intra-individual variability on the efficacy of speech-based models remains relatively unexplored. For example in \cite{schuller2014munich_short,mesleh2012heart_short,smith2017analysis_short,usman2021heart_short}, it has been shown that acoustic features, in particular spectral features, can be used as good predictors of heart activity parameters like beats per minute (BPM) in the context of regression and classification tasks. Nevertheless, the speaker-dependent data split used in these efforts might be confounding the reported performance since both training and testing sets have included data samples belonging to the same speakers. Furthermore, an interesting study conducted by \cite{schuller2013automatic_short} has addressed the speaker-dependency in their experimental setup showing the superior performance achieved using speaker-dependent split over leave-one-speaker-out (LOSO) approach. However, the level of variability between and within speakers has not been examined.

Part of the challenge in properly addressing these confounding parameters has been manifested in the lack of large speech corpora that includes recorded biological signals, leading to overfitting models on speakers' samples. One way to alleviate this limitation is to pre-train models on large scale corpora and evaluate the pre-trained models on downstream paralinguistic tasks with limited data samples. One major breakthrough in this domain is introducing self-supervised models (SSMs). These are models trained to optimize a certain task objective without using labels. Consequently, several benchmarks have evaluated the predictive power of these data-driven representations on paralinguistic tasks \cite{scheidwasser2022serab_short, shor20_interspeech_short} as well as speech \cite{yang2021superb} and audio tasks in general \cite{turian2022hear_short}. Additionally, SSMs outperformed acoustic and knowledge-based features on multiple audio-based tasks.

\begin{figure*}[ht!]
\centering
    \includegraphics[width=0.9\textwidth]{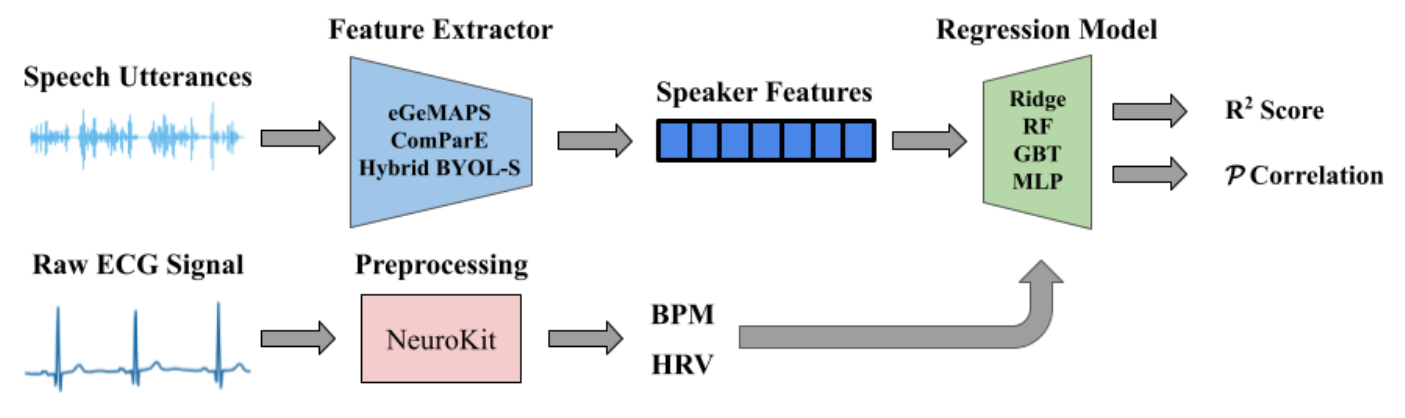}
    \caption{Training pipeline for predicting BPM and HRV values from knowledge-based and data-driven speech representations.}
\label{fig:pipeline}
\end{figure*}


In this study, we evaluate, for the first time, the ability of SSMs to be used for predicting heart activity parameters such as BPM and heart rate variability (HRV) and assess its predictive power compared to acoustic features. Importantly, we highlight the impact of both inter- and intra-individual variability on performance and generalizability. We further examine the optimal context window duration of speech for this task. Lastly, we pinpoint the salient features for this task and compare it to previous work in literature. Collectively, these analyses highlight the importance of collecting data systematically to provide better understanding of the underlying physiological variability across individuals.



The rest of the paper is organized as follows. We introduce our proposed method and the database used in Section \ref{sec:methods}. The results has been given in Section \ref{sec:results}. We conclude the paper in Section \ref{sec:discussion}.

\section{Materials \& Methods}
\label{sec:methods}

\subsection{Dataset}
\label{sec:data}
In this work, we use Ulm-TSST database presented in the MuSe challenges \cite{stappen2021muse_short, christ2022muse_short}. This corpus included 69 German-speaking subjects (49 Females and 20 Males) performing free speech tasks following the Trier Social Stress Test (TSST) protocol \cite{kirschbaum1993trier_short}. Multiple physiological signals such as ECG, BPM, respiration (RESP), and EDA were captured during the speech tasks at a sampling rate of 1 kHz. This dataset was originally used to study emotions (i.e., arousal and valence) of people in stressful dispositions. However, in this paper, we are only interested in the recorded heart activity (ECG).

\subsection{Data Preprocessing}
The utterances were originally acquired with 6 channels. We select the channel that featured the highest loudness (the first channel). Then, we re-sample all mono-channel utterances to 16 kHz and standardize them. Subsequently, we chunk the data into clips of varying window sizes ranging from 3 to 5 seconds with a hop size of 500 ms. The ECG signals are preprocessed using \texttt{NeuroKit} package \footnote{\url{https://neuropsychology.github.io/NeuroKit/functions/ecg.html}}. BPM is computed from the ECG signal using the same package. Then, for each audio clip, we extract the corresponding ECG signal and compute HRV. Lastly, we compute the average value for BPM and HRV for each clip to have one value per audio sample/clip.

\begin{figure*}[ht]
    \centering
    \includegraphics[width=\textwidth]{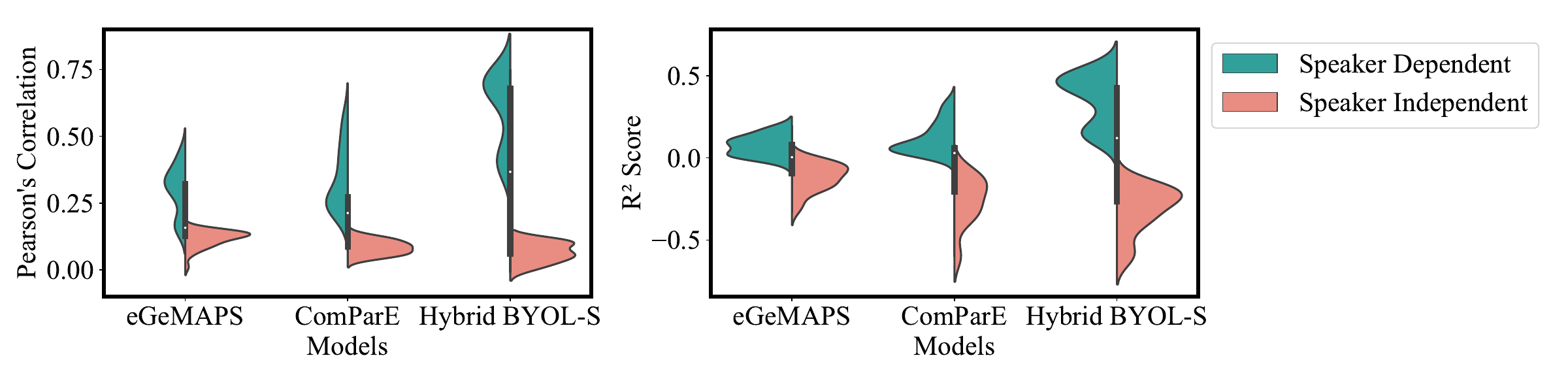}
    \caption{Performance of different speech features in both speaker conditions. The reported distributions show the evaluation across multiple regressors and window sizes as well as the performance for predicting both targets (i.e. BPM and HRV).}
\label{fig:perf_sp_cond}
\end{figure*}

\subsection{Speech Feature Extraction}
\label{sec:speech_feat}
We compared the performance of knowledge-based features against speech representations generated from a pre-trained self-supervised model. 

\textbf{\textit{Knowledge-based:}} We use openSMILE \cite{eyben2010opensmile_short}, an open source toolkit for extracting low-level descriptors from audio utterances. We extract two commonly-used feature sets from openSMILE, namely eGeMAPS (88 features) \cite{eyben2015geneva_short} and ComParE (6373 features) \cite{schuller13_interspeech_short}.


\textbf{\textit{Data-driven:}} We use Hybrid BYOL-S model \cite{elbanna22_interspeech_short}, a self-supervised model derived from Bootstrap Your Own Latent (BYOL-A) learning framework \cite{niizumi2021byol_short}. BYOL-A learns general-purpose audio representations from juxtaposing two augmented views of a single input utterance. The two augmented views are fed to two networks, an online and a target network. The task objective involves the online network predicting the generated representations from the target network. The new variant of BYOL-A (i.e. Hybrid BYOL-S) is a speech-specific derivation that trains the online network to predict data-driven representations from the target network as well as knowledge-based representations (ComParE features) simultaneously yielding robust speech embeddings of size 2048. We selected Hybrid BYOL-S due to its superior performance on multiple audio downstream tasks \cite{elbanna2022byol_short} as part of the HEAR 2021 challenge \cite{turian2022hear_short}. 



\subsection{Evaluation Pipeline}
\label{sec:eval}
We evaluate the performance of the three feature sets to predict BPM and HRV. To examine the generalizability of the methods to various speakers, we create two data splits, one speaker-independent split where we used 70\% of the speakers (N=48) for training and hold out 30\% (N=21) for testing. The second split is speaker-dependent in which we keep 70\% of the clips for each speaker in the training set and the remaining clips per speaker are evaluated. Additionally, we train speaker-specific regression models for each speaker where the model is only trained on data samples from one speaker and these samples are split into 70\% and 30\% for training and testing, respectively. For each experiment, multiple regressors are used such as Ridge regression, Random Forrest (RF), Gradient boosting tree (GBT), and multi-layer perceptron (MLP). We run a grid-search for hyperparameters that are model-specific. In case of speaker-independent, we run a 5-fold group shuffle split where speakers in the training set were further divided into train and validation sets (70\% and 30\%, respectively). Thus, ensuring that the validation set includes unseen speakers during training with cross validation. On the other hand, when running speaker-dependent experiments, we use a 5-fold time series split where the training samples for all speakers are split into train and validation (70\% and 30\%, respectively) while considering the temporal dependency between the data samples. Lastly, for speaker-specific, we train regression models on data samples from a single speaker using 5-fold time series split as well. After training the regression models, we report performance on a held out test set using coefficient of determination ($R^2$) score as well as Pearson's correlation coefficient, as illustrated in Figure \ref{fig:pipeline}.

\section{Results \& Analysis}
\label{sec:results}
Figure \ref{fig:perf_sp_cond} illustrates the performance of candidate speech representations on predicting BPM and HRV under two different speaker conditions for unseen test sets; speaker-dependent and speaker-independent protocols. We report the $R^2$ and Pearson's correlation between the predicted and the ground truth. The plotted distributions highlight the performance across different regressors, targets (i.e., BPM and HRV), and window sizes (i.e., 3, 4, and 5 seconds). We show that Hybrid BYOL-S outperforms knowledge-based features in a speaker-dependent setting. Whereas, all representations perform equally poorly in the speaker-independent setting. This result highlight the limitations of generalizability of speech features for this downstream task.

\begin{figure}[ht]
    \includegraphics[width=0.5\textwidth]{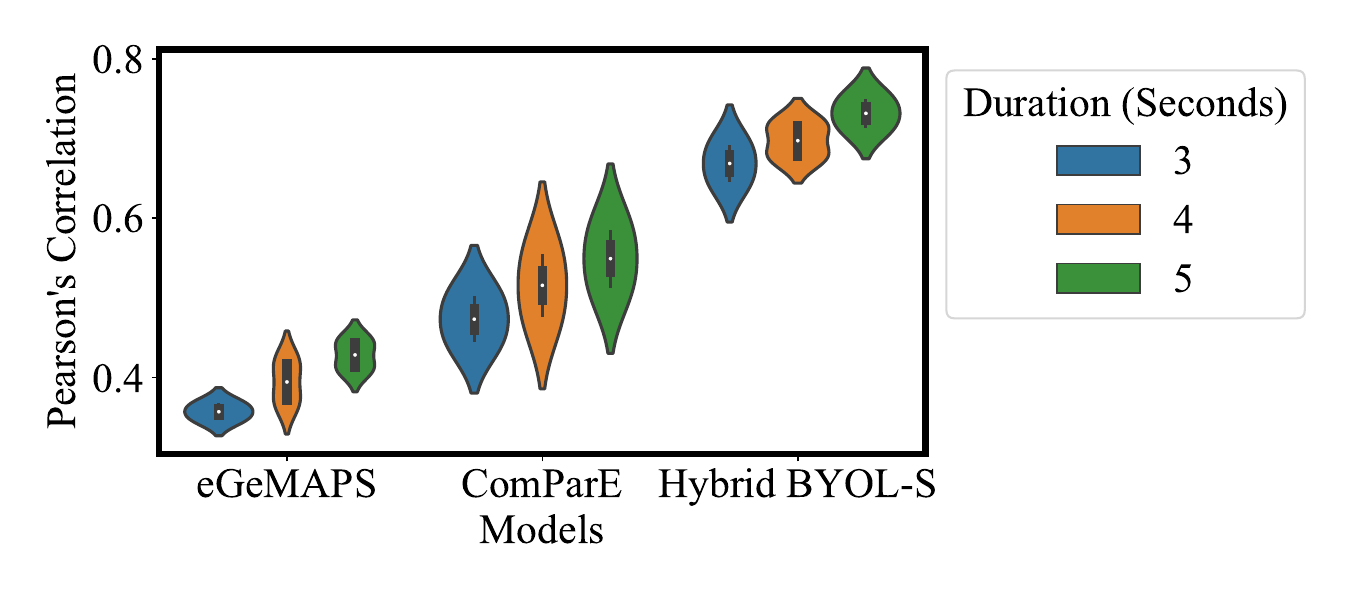}
    \vspace{-1cm}
    \caption{Performance of different speech features with varying context window duration. The reported distributions show the evaluation using speaker-dependent and GBT regression model for predicting both targets (i.e., BPM and HRV).}
\label{fig:win_size}
\end{figure}

Moreover, we study the effect of context window duration on performance. Figure \ref{fig:win_size} shows that increasing the window size improves performance in all feature candidates. This figure reports the Pearson's correlation on a speaker-dependent test set using GBT (best-performing regressor) for both targets. Importantly, we observe significant improvement between 3 sec window and 4 sec window while between 4 sec and 5 sec duration performance is relatively overlapping.

\begin{figure}[!htb]
    \centering
   \includegraphics[width=0.5\textwidth]{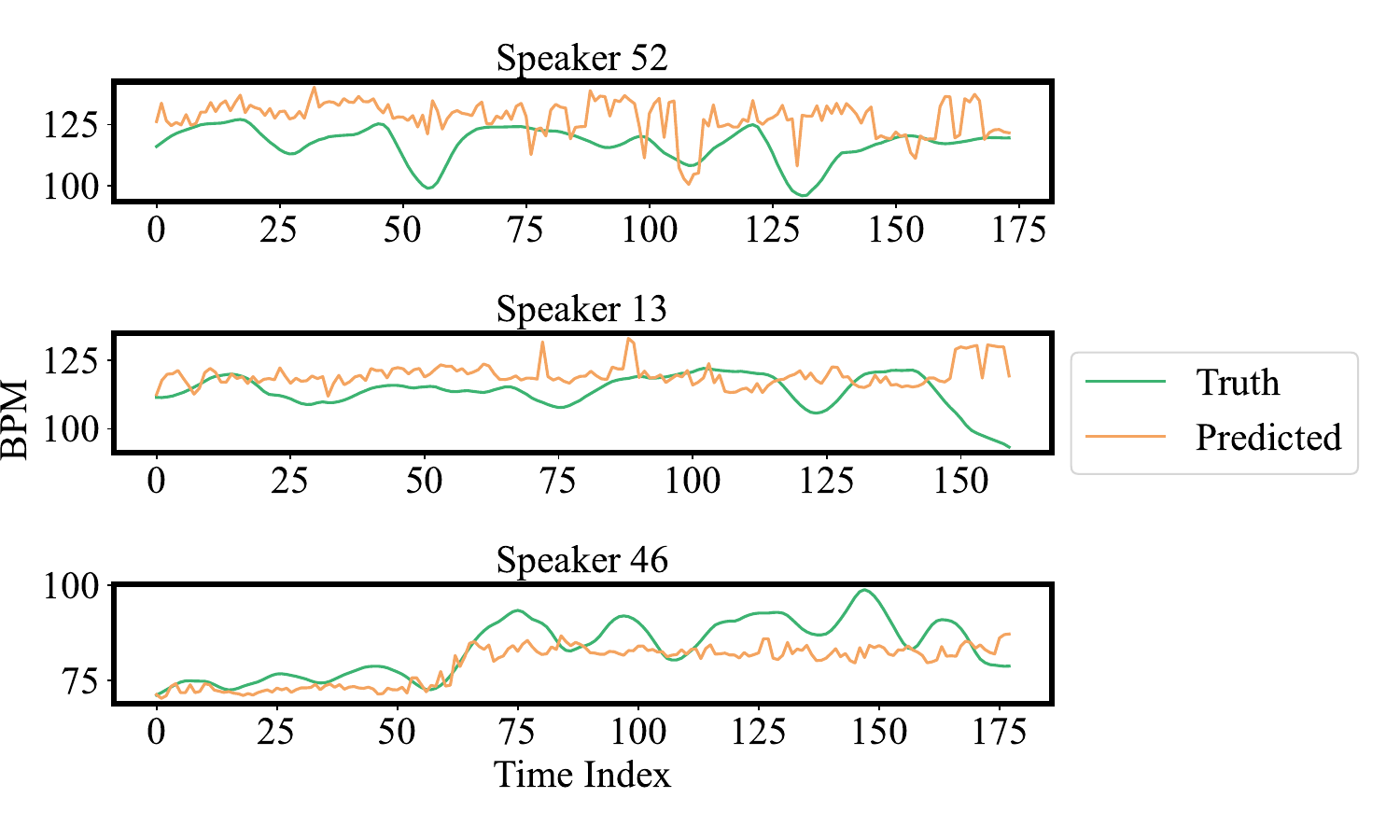}
   \vspace{-0.5cm}
   \caption{Predictions from GBT model using Hybrid BYOL-S features with 5 sec window size. Predictions are shown for speakers 52, 13, and 46, respectively.}
\label{fig:sp_pred}
\end{figure}

\begin{figure*}[ht]
    \includegraphics[width=\textwidth]{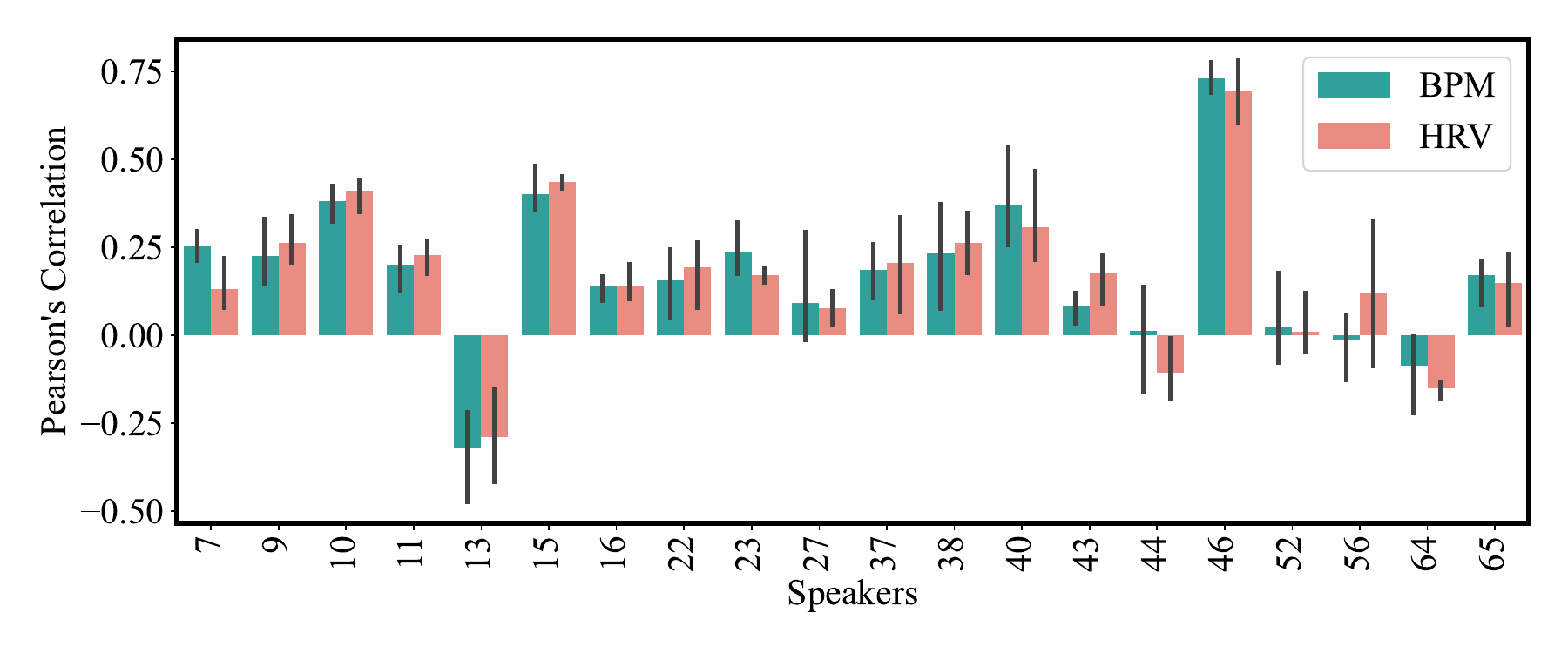}
    \vspace{-1cm}
    \caption{Performance of Hybrid BYOL-S features for speaker-specific protocol across all window sizes (3, 4, 5 seconds) of audio and a GBT regressor model. The figure shows a random sample of 20 speakers for both targets BPM and HRV}
\label{fig:sp_specific}
\end{figure*}

Given the discrepancy between both speaker-related protocols, we further train regression models on data samples from a single speaker resulting in building a model per speaker (i.e., speaker-specific protocol). Figure \ref{fig:sp_specific} demonstrates the high variability in performance across a randomly selected sample of 20 speakers. In this figure, we report the Pearson's correlation evaluated on Hybrid BYOL-S to predict BPM and HRV across different window sizes.

We further plot the predictions from three speakers; the speaker exhibiting the lowest correlation (i.e., speaker 52), the speaker with the highest positive correlation (i.e., speaker 46) and the speaker with the highest negative correlation (i.e., speaker 13), as shown in Figure \ref{fig:sp_pred}. This figure provides insights regarding the model's ability to capture the variability within a single speaker.

\begin{figure*}[!htb]
    \includegraphics[width=\textwidth]{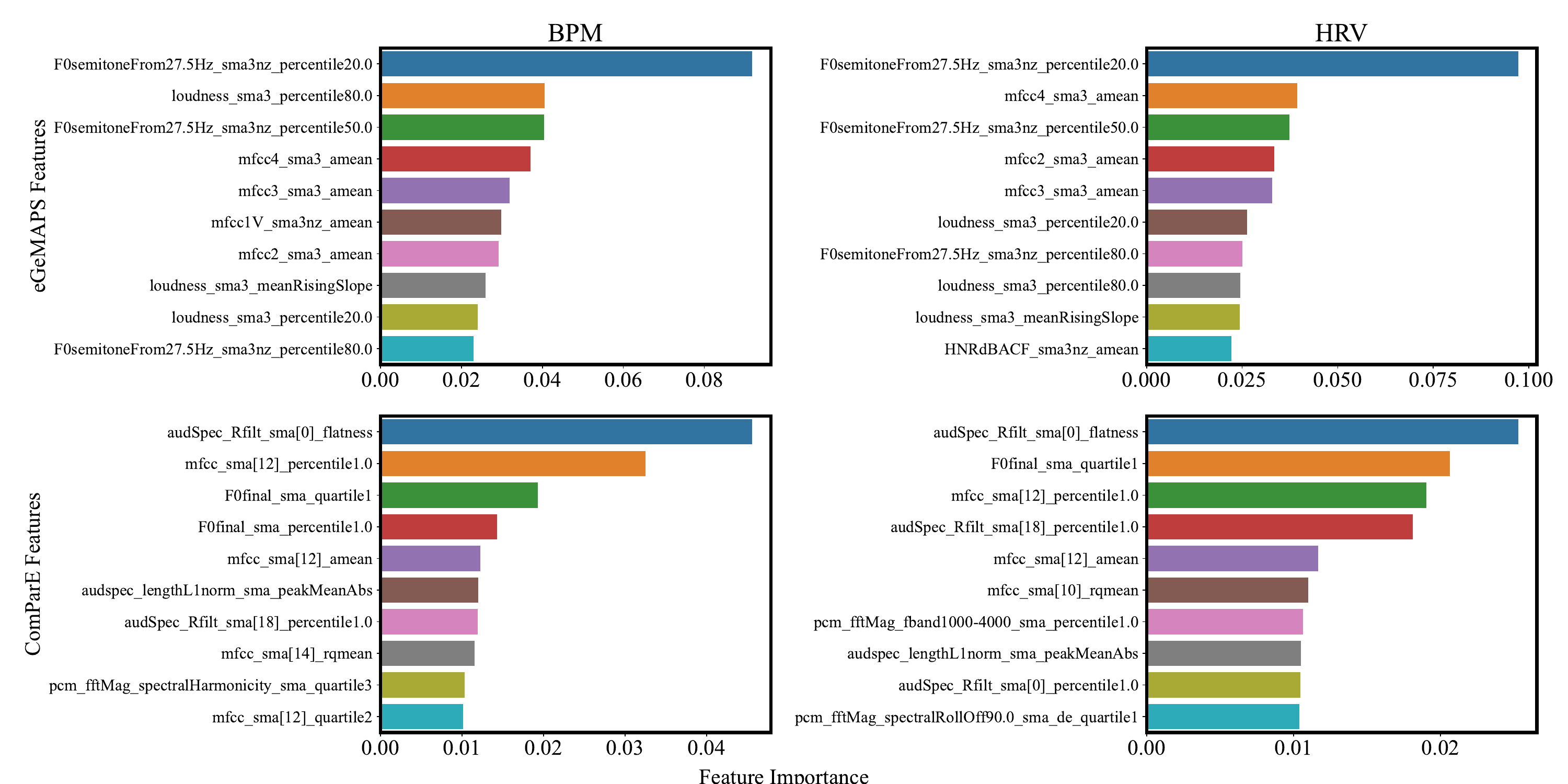}
    \caption{Feature importance showing top 10 acoustic features for both targets BPM and HRV extracted from both openSMILE feature sets eGeMAPS and ComParE.}
\label{fig:feat_imp}
\end{figure*}

Furthermore, we opted to pinpoint the salient acoustic features for predicting heart activity. Thus, we perform feature importance on the trained GBT regression model using eGeMAPS and ComParE features. Figure \ref{fig:feat_imp} shows the top 10 features for each feature set that contribute to predicting both BPM and HRV. The feature importance is computed from the best estimator and ranked accordingly.

\section{Discussion \& Conclusions}
\label{sec:discussion}
In this paper, we studied the robustness of acoustic and data-driven features to predict heart activity parameters. We showed that Hybrid BYOL-S, a self-supervised model, outperformed acoustic features on this downstream task. This finding aligns with previous work showing the feasibility of using speech representations as predictors for heart activity \cite{schuller2014munich_short,usman2021heart_short}. Additionally, we highlight the significance of leveraging data-driven, in particular SSMs, representational spaces for better predictive power. 

Nevertheless, we noticed that all features performed equally poorly in the speaker-independent protocol as shown in Figure \ref{fig:perf_sp_cond}. This result exposes the limitations of these features and their sensitivity to inter-individual variability confirming the work done in \cite{schuller2013automatic_short}. Importantly, this dataset was originally collected to study speech as a biomarker for stress. Accordingly, one might hypothesize that the high variability in heart activity might have been a consequence of variability in stress response across individuals. Hence, the inability of models to generalize across speakers. Thus, it is crucial for future work to consider emotional states and cognitive aspects as confounding variables for predicting heart activity from speech. We further observe varying correlation scores when training models on a single speaker data as in Figure \ref{fig:sp_specific}. The poor performance in some speakers indicates that even intra-individual variability plays a role as well in hindering the model to properly predict unseen data samples from the same speaker which has not been previously discussed. Moreover, we demonstrated that increasing the context window duration more than 3 sec provides significantly better predictions.

Lastly, we ran feature importance as shown in Figure \ref{fig:feat_imp} and found that spectral features and loudness were the most important acoustic features for this downstream task which line up with the previous findings in literature \cite{orlikoff1989effect, mesleh2012heart_short, schuller2014munich_short}. These results together provide insights on the robustness and limitations of acoustic and data-driven features in studying heart activity as well as shed light on the importance of adding the mental state as an essential dimension to study the relationship between speech and physiological signals such as heart activity.

\section{Acknowledgments}
This work was partially funded by the Swiss National Science Foundation through the project Towards Integrated processing of Physiological and Speech signals (TIPS), grant no. 200021\_188754. The first author carried out this work as an intern at the Idiap Research Institute, Martigny, Switzerland.

\newpage

\bibliographystyle{IEEEtran}
\bibliography{mybib}

\end{document}